\documentclass[a4paper]{article}

\usepackage{graphicx}

\usepackage{hyperref} 

\usepackage{amsmath}
\usepackage{amssymb}

\usepackage[normalem]{ulem}

\usepackage{epstopdf} 

\usepackage{alltt}

\begin{document}

\title{Concept-Oriented Model: the Functional View}

\author{Alexandr Savinov \\ \href{http://conceptoriented.org/savinov}{http://conceptoriented.org/savinov} }

\date{06.06.2016}

\maketitle

\begin{abstract}
The plethora of existing data models and specific data modeling techniques is not only confusing but leads to complex, eclectic and inefficient designs of systems for data management and analytics. The main goal of this paper is to describe a unified approach to data modeling, called the concept-oriented model (COM), by using functions as a basis for its formalization. COM tries to answer the question what is data and to rethink basic assumptions underlying this and related notions. Its main goal is to unify major existing views on data (generality), using only a few main notions (simplicity) which are very close to how data is used in real life (naturalness).
\end{abstract}

\section{Introduction}
\label{intro}

\subsection{Technological Trends and Challenges}
\label{1.1}

\textit{Data} is fuel and prerequisite for any kind of data management, analysis and decision making process. However, with the explosion of data volume and the variety of data sources --- two aspects of the big data problem \cite{Coh09} --- we observe quite significant difficulties in applying conventional methodologies to real world problems. The existing theories and technologies for data management and analytics have been pushed to the limits of their ability to solve more and more complex tasks especially in the context of significant modern trends over the last few years which are shortly described below. 

\textit{Agile analytics} goes beyond standard OLAP analysis by facilitating ad-hoc queries where the user can freely vary data processing and/or visualization parameters and is not restricted by predefined application-specific and domain-specific scenarios. 

\textit{Self-service analytics} is one of the most significant trends in the BI industry over the last few years. It is aimed at giving users the ability to solve analytical tasks with little or no help from IT \cite{Sav14a}. Self-service tools are intended for such users as data enthusiasts, business users, data artisans, analysts. What unites all these kinds of users is that they do not possess deep knowledge in mathematics and statistics but all of them need some simple tool to solve a problem or answer a question by analyzing available data.

\textit{Near real time analytics.} There is strong demand in reducing the time between data acquisition and making a business decision but conventional systems cannot provide the necessary response time and agility of decision making on large volumes of data \cite{Cha11,Thi12}. Although modern hardware provides a basis for a new generation of in-memory, columnar databases \cite{Man00} with potentially higher query performance on analytical workloads, it is important to understand that it is not only a hardware problem -- new data models, query languages, analysis scenarios and algorithms are needed. 

\textit{Data semantics.} A typical enterprise system can contain tens of thousands data tables and open systems can involve even more external data sources. In this situation it is extremely difficult to get meaningful results without understanding what data means and having the possibility to process data automatically \cite{Sto93,Dah05}. In this context, semantics ``should enable to respond to queries and other transactions in a more intelligent manner'' \cite{Cod79}. 

\textit{Reasoning about data.} Analytical queries are rather complex data processing scripts over numerous data sources and writing such queries is a tedious and error-prone task requiring high expertise. The mechanism of reasoning about data can significantly simplify this problem by automatically deriving the desired result from the available data. 

\textit{Analytical computations.} Analysts need to embed complex computations in their analysis tasks and normally it is performed by copying data to a different system for processing which is both inefficient and not flexible procedure. Executing custom analysis tasks close to the data is still a big problem because of serious incompatibilities between data modeling and programming. 

\subsection{Theoretical Issues}
\label{1.2}

Obviously it is difficult to solve these and other practical data management problems without clear understanding of what data is. However, there exist some fundamental issues and controversies which do not allow us to answer this question and hence to solve these problems in a principled manner. Some of these basic issues are shortly described below. 

\textit{Tuples.} Tuple is probably the most wide spread way to formally represent data by combining several simpler data elements. This mathematical construct has been used for representing complex things in computer science for dozen of years and it seems that everybody is happy with its properties. Yet, one serious issue arises (at least in the context of data modeling) if we ask the question whether a tuple is a value (passed by-copy) or it is passed by-reference. If a tuple is a value then the whole database is essentially one huge tuple because there is no possibility to reference things. If tuples are passed by-reference then any tuple is a combination of references and there is no way to represent a combination of values. In both cases, something very important is missing when using tuples as a mathematical construct for data representation. Therefore, in practice there exist two major workarounds: 1) two separate structures are used for representing values and objects (either as dedicated structures like struct and class or by annotating them by a marker like by-ref and by-val); 2) each individual instance (variable, parameter, field and so on) is annotated as passed by-value or by-reference. In particular, a variation of the first approach is used in the relational model (RM) \cite{Cod70} where two kinds of data elements and two kinds of sets are used: values are members of domains and tuples are members of relations. A fundamental controversy here is that both domains and relations are normal sets containing tuples and therefore there is no formal reason to distinguish them. So the question is why \textit{two} kinds of sets for representing elements with the same structure (tuples)? Why tuples from relations can be composed of only tuples from domains and not from tuples from other relations? Why tuples from domains cannot include tuples from relations? Of course, there exist numerous specific fixes and workarounds for solving these problems but the question is about inherent support at the level of tuples themselves as a basic construct for data representation and access. And the question is then whether it is possible to introduce \textit{one} kind of tuples for both purposes? 

\textit{Sets.} A set is a fundamental mathematical notion which is widely used in data modeling for formally representing a collection of distinct things. Yet, there exist several subtle questions which have different answers in different models. One of them is whether a set itself is a data element or only its members are data elements? In real life, a collection of objects is normally treated as a new object. In RM, a set (relation or domain) is not a fully-fledged data element. Other approaches like object-based models introduce collections which can be treated as data elements (but also with some limitations and specific treatments). Another question is whether every data element should be a set? For example, in RM a tuple is not a set. What is the difference between tuples and sets if tuples in mathematics are defined via sets? Is this difference important for data modeling or maybe it is enough to have only sets and not tuples? Should any data element (including sets) be a member of some other set and whether it is possible to have data elements without a set it is in? If all elements including sets are included in other sets then all of them should be members in some kind of global set. How this global set should be organized? 

\textit{Hierarchies.} A hierarchy is one of the most natural ways to organize things and to think about the world (along with tuples which combine elements). Yet, the use of hierarchies has always been controversial in data modeling: some techniques (like XML) and methodologies like object-based models \cite{Dit86,Atk89} provide full support for hierarchies while others like RM essentially expel them from the data realm. One reason for not supporting hierarchies is the existence of multiple treatments of their meaning. Does a hierarchy represent containment? Is this containment in terms of membership or subset relations? And if it is containment what is the difference from tuples regarded as containers for the members? Or maybe a hierarchy represents inheritance? Then what mechanism is used for the implementation of inheritance? Or maybe it corresponds to a dimension hierarchy? Due to this diversity of interpretations, systems normally provide different hierarchies for different purposes while a hierarchy as a basic construct loses its role (and it is probably one of the reasons why they are not supported in RM). Another more specific but rather deep issue is asymmetry between class or concept hierarchies and the corresponding instance hierarchies. The paradox \cite{Ste87} is that instances forget about the hierarchical type structure of their classes and exist in a flat space rather than as a hierarchy. A general question is whether it is possible to have only one kind of hierarchies as a basic construct and at the same time cover all their common uses and interpretations? 

\textit{Multidimensionality.} Probably nobody after Descartes will dispute that the world is inherently multidimensional and multidimensionality is one of the most common ways to organize things (along with tuples and hierarchies). Indeed, it is rather natural to assume that similar to objects in physical space, data items and conceptual things also have coordinates which determine their position with respect to other things. Surprisingly, inspite of its naturalness and importance for data analysis, multidimensionality is not an inherent feature of general-purpose data models. Rather, multidimensionality is added as a new layer of data representation for (mostly numeric) analysis purposes \cite{Ped09,Ped01}. Therefore, the question is whether it is possible to have multidimensionality as an inherent mechanism of the model at the same level as tuples, sets and hierarchies so that data always exist within a multidimensional space? 

\textit{Identification.} Identity determines what a thing is, how things exist, how they are distinguished, how they are accessed as well as many other mechanisms \cite{Ken03}. Currently there exist many different approaches to identification of elements like pointers, references, surrogates, keys, oids or identification by content. Here again there exist many quite different views on the role and importance of identities for data modeling. In particular, there are different opinions concerning the question whether identities are data at all and if yes then whether identities belong to the problem domain or they belong to the system level. Finding some common basis for all of the existing views on identities and identification would significantly simplify many complex data modeling and analysis problems. 

\textit{Connectivity.} All things in the worlds are connected and all elements in a database have to be also connected. One way to connect things is via properties storing a reference to a related thing. Another approach, joins, is where two things are considered related if they both contain some common value. And the third approach consists in using relationships which are elements referencing related things. Do we actually need all these mechanisms? And if yes then is it possible to introduce one basic principle all other connectivity mechanisms can be reduced to? 

\subsection{Goals and Contribution}
\label{1.3}

These fundamental challenges can hardly be resolved by introducing a new specific method or technique. They require a principled solution which should be general enough to unify many existing patterns of thought, and at the same time it should be simple and natural. Such unification is the main goal of the concept-oriented model (COM) described in this paper. COM addresses the theoretical and practical issues described in the beginning of the paper and is aimed at \textit{simplifying} data modeling and management. Of course, complete unification is hardly possible because it would mean developing a kind of the ultimate theory. Therefore, COM should be viewed as an attempt to increase unification among existing theories of data and data modeling techniques. COM tries to achieve higher \textit{generality} by decreasing the number of basic notions used to describe a database by simultaneously increasing their coverage, that is, the number of various data modeling patterns that can be effectively modeled by these notions. Also, COM is being developed to be a \textit{natural} approach so that its main notions are close to how data is thought of and used in real life (semantics). 

COM has been described at conceptual level as well as syntactically using the concept-oriented query language (COQL) \cite{Sav11a,Sav12d,Sav14b} with limited formalization. COM has also been implemented in two systems: a self-service tool for analytical data integration, ConceptMix \cite{Sav14a} and a framework for data wrangling and agile data transformations, DataCommandr \cite{Sav16a}. The main contribution of this paper is that we propose a new formalization of COM in terms of functions, sets and tuples. More specifically, we make the following contributions:

\begin{itemize} 
\item New formal definition of a data element via two kinds of functions: identity functions and entity functions is proposed. 
\item New mechanism of function overriding and data hierarchies is described. 
\item New use of functions for data modeling is described where functions are mappings between values with arbitrary domain-specific structure (rather than only primitive identifiers). 
\item New treatment of connectivity based on functions where joins and relationships are considered two higher levels types of connectivity derived from functions. 
\item New uses of functions for representing data semantics: a function is a mapping from an object to its coordinate, a function is a mapping from more specific elements to more general elements, a function is a mapping from members to their set. 
\end{itemize}

The paper has the following layout. In Section 2 we introduce basic but the most important notions of data element, function and concept. Section 3 defines the structure of data elements which is a nested partially ordered set. Section 4 describes the structure of sets which is opposed to the structure of elements. Section 5 is devoted to describing how the formal constructs of COM can be interpreted. Section 6 provides a discussion of this approach, its benefits and subtle properties. In Section 7 we describe how the principles of COM are used in a novel column-oriented framework for agile data integration and transformations, and Section 8 makes concluding remarks.

\section{Data}
\label{data}

\subsection{The Value of Values}
\label{2.1}
\textit{Values.} COM makes a very simple and natural assumption that a data element is a \textit{value}. In computer science, the main characteristic of values is connected with the mechanism of representation: values represent themselves directly by their content and can be passed exclusively by copying the whole content. In particular, it is not possible to share a value or to represent it indirectly via another value (if it is not part of some other construct as will be described below). Anything can be considered a value if its contents can be copied as one whole. Examples of values are numbers like 25.76 or letters like `b'. 

\textit{Copies and locations.} If one value can produce many copies then an important question arises: is a copy of a value equivalent to the original value or it is a different element with the same content? For example, if number 5 is copied from one database to another then will the second copy be treated as a new element or there still exists only one value? It is a quite controversial question because both options are important. On one hand, there is only one number 5 independent of where and how it is used. On the other hand, copying a value is a creation procedure which by definition produces something new and we need to take it into account. If values are considered in isolation then COM assumes that all copies of one value are equivalent (as it is accepted in mathematics). However, multiple copies of one value have different locations. These locations provide a context which allows us to distinguish different representations of the same data. Thus COM distinguishes between values themselves and their existence in different locations. When we say that a value was changed then we actually mean that the content in some location was changed and it does not influence other locations. 

\textit{Tuples.} Any value has some structure. Values the structure of which is hidden or ignored are referred to as \textit{primitive values} (also called elementary, system or atomic values in other models). New values can be created by using a \textit{composition} of existing values. Formally, such a composition is represented by \textit{tuples} which are treated in their accepted mathematics sense by capturing the notion of an ordered list. A tuple is a value written in angle brackets  $\langle \ldots \rangle$  enclosing its member values. A tuple consisting of $n$ members is called $n$-tuple and $n$ is called its \textit{arity}. Tuple members are distinguished by their relative location which in COM is referred to as \textit{dimension} (also called attribute, column, property, slot, variable or characteristic in other models). Dimension and the corresponding member are separated by colon (we also used equality in other publications). If  $e = \langle \ldots,f:a,\ldots \rangle$  is a tuple then $f$ is a dimension and  $a$  is a member stored in this dimension. By definition (of values), tuples are composed of copies of the member values which get distinct locations in the tuple. Thus tuples are constructs where copies of other values exist. But since a tuple itself is also a value, it cannot be shared or represented indirectly -- it can be only copied, for example, become a member of another tuple. There is one special value represented by empty tuple  $\langle\rangle$  (also denoted by NULL) which has empty structure. It is assumed that adding empty value to a tuple (in any dimension) does not change it:  $e = \langle a_1,a_2,\ldots,a_n, \langle\rangle\rangle = \langle a_1,a_2,\ldots,a_n \rangle$. To avoid infinite structures we prohibit inclusion of a (non-empty) value into itself directly or indirectly and hence value composition is an acyclic graph. 

\textit{Extension} is an approach to building new values by using composition (tuples) where one constituent (tuple member), called \textit{super-element} or base element, has a special semantic role and uses. All other constituents taken as a whole are referred to as an \textit{extension} or \textit{segment}, that is, a segment or extension is a tuple without its base. We will use a convention that a base dimension has a special name \textit{super} and is written as the first member separated from the extension members by bar symbol: $e = \langle super:b | \ldots, f:a, \ldots\rangle$. Here  $b$  is a super-element and  $a$  is a member of the extension. It can be also written as  $e = \langle b | a \rangle$  where  $b$  is a base and  $a$  is an extension. A tuple with empty extension is equal to the base. A tuple with empty base is equal to the extension:  $\langle \langle\rangle | e \rangle = e$. Therefore, \textit{any} tuple can be represented as a sequence of segments where the very first segment is empty tuple and each next segment extends the previous one: $e = \langle super:\langle super: \langle\rangle | a \rangle | b \rangle | \ldots \rangle$  where $a,b,\ldots$ are segments. All possible tuples are then represented as a tree of extensions where empty tuple is a root, a parent is a base for its children and a child is an extension of its base. It is a tree because there is only one base for any tuple and cycles are prohibited. Note that in this definition of extension, there is no difference from normal tuples except that one dimension has a special name (\textit{super}). In the next section we will describe how this special dimension is used to constrain possible tuples. 

\subsection{The Function of Functions}
\label{2.2}

\textit{Functions} in COM are treated in their accepted mathematical sense as a mapping from a set of input values into a set of output values where exactly one output value is associated with each input value. A function is denoted as  $f(x):D \rightarrow R$  where $D$ is a set of input values, called \textit{domain}, $R$ is a set of all output values, called \textit{range}, and $x$ is an argument that takes its values from $D$. Function input can be written either as an argument in parentheses,  $y=f(x)$, or before the function using \textit{dot notion},  $y=x.f$. One element of the mapping defined by a function  $\langle x,y=f(x) \rangle$  is referred to as a \textit{function element}. Thus a function can be represented as a set of all its elements:  $f=\{ \langle x,f(x) \rangle \}$, $x \in D$. One distinguishing feature of COM functions is that formally they define some output for \textit{all} input values. However, the empty output, $\langle\rangle=NULL$, is semantically interpreted as having no output, that is, the input is not mapped. If all inputs from the domain are mapped to empty value, $\forall x \in D$, $f(x)=\langle\rangle$, then this function is called \textit{empty function}. The benefit is that we can formally work with functions returning empty values precisely as with normal functions by forgetting about the fact that not all input values can be mapped. From the data modeling point of view, a distinguishing feature of COM functions is that they define mappings between values with \textit{arbitrary} domain-specific structure rather than only between primitive types. 

\textit{Identity and entity functions}. In the previous section we defined value as a tuple composed of other values with certain positions. An alternative representation of composition is that tuple dimensions are treated as functions (rather than locations or positions) which return tuple members: if  $e=\langle\ldots,f:a,\ldots\rangle$  then  $f$  is a function such that  $a=f(e)$  or  $a=e.f$  in dot notation. Then a value is represented as a tuple of functions and these member functions are used to access tuple members given a tuple. Now let us ask a question: what happens with these functions if a tuple is copied? A function is a mapping that can be represented as a set of function elements, and hence there are two possibilities: elements of this mapping are also copied or the mapping is shared among all copies of the tuple. According to this distinction, a function which is passed by-value by copying its definition (mapping) is referred to as an \textit{identity function} and a function which is passed by-reference is referred to as an \textit{entity function}. It is convenient to think of an identity function as storing its outputs within the tuple itself (so a tuple is used as storage) and hence identity functions are copied along with tuples. In contrast, entity functions store their outputs outside of the tuple and there exists only one such definition shared among all tuples. 

\textit{Data element.} COM uses two kinds of tuples: \textit{identity tuples} are composed of only identity function outputs and are written in angle brackets  $\langle\ldots\rangle$. \textit{Entity tuples} are composed of only entity function outputs and are written in parentheses  $(\ldots)$. A \textit{data element} in COM is defined as an identity tuple with an associated entity tuple the entity functions of which are defined on the identity: 
\begin{equation} 
\mbox{[data element] }
e=\langle \ldots,f:a,\ldots \rangle(\ldots,g:b,\ldots)
\end{equation}

\noindent
Importantly, only identity is a value while entity tuple (outputs of all entity functions) is not a value. An entity cannot be copied or passed as one whole. Only entity constituents -- outputs of individual entity functions -- are values and can be copied. If identities are normal values then why do we call them identities? Because identities are values with an associated entity and identities provide the only way to represent and access entities. Thus identities represent not only themselves (by-value) but also the associated entity (by-reference). Two elements are considered identical if they have the same identity. 

\begin{figure}
\begin{center}
\includegraphics[width=80mm]{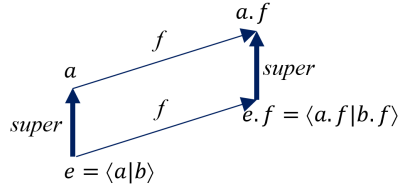}
\caption{Function extension principle} \label{figure01}
\end{center}
\end{figure}

\textit{Function extension.} Generally, a tuple can involve functions with the same name as in its members but they will be completely unrelated. Yet, it is not so for super-elements and it is one of the main reasons for their introduction. If an extension defines a function already used in its super-element, $e=\langle super:\langle \ldots,f:a,\ldots\rangle | \ldots,f:b,\ldots\rangle$, then the extension is said to \textit{override} the super-function. COM introduces a novel mechanism for overriding functions. Instead of completely hiding the super-function and returning an arbitrary value, an extension makes a contribution to the value returned by the super-function. In the above tuple, the value  $a=f(e.super)$  returned by the super-function cannot be completely overridden by the extended function but rather it is modified. Formally, overridden functions must satisfy a function extension principle which postulates that \textit{a value returned by a function must extend the value returned by its base function} (Fig.~\ref{figure01}): 
\begin{equation} 
\mbox{[function extension] }
e=\langle a|b \rangle \Rightarrow e.f=\langle a.f|b.f \rangle
\end{equation}

Function output can be represented as a sequence of values each returned by this function applied to one segment of the identity. If a function is not defined for a segment then its output is supposed to be empty tuple which does not change the result. The function extension principle can be expressed in terms of partial order as will be described in Section 3.2. Its novelty and main benefit is that data fields can be overridden by making them more specific in extensions. An example can be found in Section 4.2 and in \cite{Sav12c} where also the reverse overriding strategy is described. 

\subsection{The Concept of Concepts}
\label{2.3}

\textit{Concept} is a syntactic construct for describing the structure of data elements in COM. It is a template which only declares functions but does not provide their definitions (it does not define function elements). Concepts are used as data types for various storage elements like variables, fields, parameters, collections etc. A concept is defined as a couple of two classes: one identity class and one entity class. Both classes are defined as a combination of functions (also called dimensions or fields) which are typed by other concepts or primitive types. If a field belongs to the identity class then it declares an identity function and if it belongs to the entity class then it defines an entity function. A concept with the empty entity class is equivalent to conventional constructs describing value types like classes in C++ or struct. A concept with the empty identity class is equivalent to a conventional class like that in Java. 

\textit{Concept instances.} Instances of identity classes are values which can be copied or stored in variables having this (or more general) concept. Instances of entity class are entities the dimensions of which can be retrieved given the identity. For example, a bank account could be described by the following concept: 

\begin{alltt} 
\textbf{CONCEPT} Account 
  \textbf{IDENTITY} 
    CHAR(10) bankNo 
    CHAR(10) accNo 
  \textbf{ENTITY} 
    Person owner 
    DOUBLE balance
\end{alltt}

\noindent 
This concept has 4 functions: 2 identity functions and 2 entity functions. Any variable of this concept will store a value with the structure described by the identity class of concept \texttt{Account} (two fields in this example). And this variable can be used to retrieve values from the entity class fields like the account balance. Thus concepts can be thought of as conventional value types describing values (say, domains in RM) with a number of associated functions returning values depending on this value. 

\textit{Concept inclusion.} A concept can be extended by another concept and this relation among concepts is \textit{called inclusion}. For example, we could define a concept \texttt{Bank} and then extend it by a concept \texttt{Account}: 

\begin{alltt}
\textbf{CONCEPT} Bank 
  \textbf{IDENTITY} 
    CHAR(10) bankNo 
  \textbf{ENTITY} 
    CHAR(10) name
\end{alltt}

\begin{alltt}
\textbf{CONCEPT} Account \textbf{IN} Bank 
  \textbf{IDENTITY} 
    \sout{CHAR(10) bankNo} 
    CHAR(10) accNo 
  \textbf{ENTITY} 
    Person owner 
    DOUBLE balance
\end{alltt}

\noindent 
Note that we removed the \texttt{bankNo} field from the \texttt{Account} concept because it is modeled via inclusion relation (extension). In other words, now the bank number field is a super-dimension. Variables of the \texttt{Account} concept will still store values with two fields: account number and bank number inherited from the super-concept. 

\textit{Overriding fields.} A sub-concept can override a field of its super-concept by making its type more specific according to the function extension principle (2) (see also type constraint in Section 3.2). For example, if super-concept \texttt{A} has a field \texttt{location} of type \texttt{Country} 

\begin{alltt}
\textbf{CONCEPT} A 
  \textbf{ENTITY} 
    Country location // Base field
\end{alltt}

\noindent 
then the sub-concept \texttt{B} can override this field by defining it as having type \texttt{City} 

\begin{alltt}
\textbf{CONCEPT} B \textbf{IN} A 
  \textbf{ENTITY} 
    City location // Overridden field
\end{alltt}

\noindent 
Importantly, \texttt{City} must be included in \texttt{Country}, that is, a field type can be made only more specific than that in the super-concept. What the overridden field will store? Instances of \texttt{B} will store only the city segment while the country segment will be stored in the super-element of type \texttt{A}. Here it is also important that one super-element can be shared among many sub-elements which means that each sub-element adds also its own extensions to the fields defined in the super-element (see also discussion of the type constraint in Section 3.2). 

\textit{Properties of concepts.} One important methodological consequence of using concepts as data types is that data modeling starts from defining values which can be used by themselves or as references (identifiers). And after that entity properties can be added as functions of identities. Identities can exist without entities but entities cannot exist without identities. Therefore data modeling is getting more value-oriented and identity-oriented in comparison to the traditional entity-focused approaches. Another consequence is that hierarchies are integral and natural part of any data type and the whole data modeling process. The main advantage of inclusion is that it automates what in many situations has to be done manually. Also, concepts generalize conventional classes and can be used as normal classes but at the same time they can model hierarchies of objects (similar to prototype-based programming and the hierarchical data model). 

\section{Structure}
\label{structure}

\subsection{Partial Order}
\label{3.1}

Having some constraints imposed on the structure of a set of data elements is important because they are assumed to be used for representing data semantics by excluding meaningless states. In mathematics, a structure imposed on a set (for example, graphs, topologies, orders, geometries or matroids) is represented by an associated relation. A specific feature of COM is that it uses partially ordered sets as a structural constraint imposed on data elements. This constraint does not depend on the kind of functions (identity or entity) and therefore in this section we will assume that an element is a conventional tuple. 

\textit{Strict partial order} is a binary relation $<$ (less than) on elements of the set  $R=\{a,b,c,\ldots\}$. If $a<b$  ($a$ is less than $b$) then $a$ is a \textit{lesser} element and $b$ is a \textit{greater} element. This relation satisfies the properties of irreflexibity and transitivity:
\[
\mbox{[irreflexivity] } \forall a \in R, \neg (a<a) \\
\]
\[
\mbox{[transitivity] } \forall a,b,c \in R,(a<b) \land (b<c) \Rightarrow a<c
\]

\noindent 
The property of antisymmetry holds as a consequence of the above two properties: 
\[
\mbox{[antisymmetry] } \forall a,b \in R, a<b \Rightarrow \neg (b<a)
\]

\textit{Partially ordered set} (poset)  $\langle R,< \rangle$  is a set $R$ with a strict partial order relation $<$ established on its elements. Element $a$ is said to be immediately less than $b$,  $a<^1 b$, if  $a<b$  and  $\nexists c:a<c<b$. The number of immediate greater elements of this element is referred to as \textit{arity} or dimensionality. \textit{Primitive} elements do not have any greater elements. 

\textit{Lattice.} If any two elements  $a,b\in R$  both have a least upper bound  $\sup⁡(a,b)$  (supremum) and a greatest lower bound  $\inf(a,b)$  (infimum) then this poset is a \textit{lattice}. For a finite set, a lattice has two special elements. The greatest element $\top$, called \textit{top}, is greater than any other element of the set:  $\forall a \in R, a<\top$. The least element $\bot$, called \textit{bottom}, is less than any other element of the set:  $\forall a \in R, \bot<a$. A \textit{labeled poset} (and lattice) is a set where all instances of the partial order relation have labels. If $a$ is less then $b$ with label  $f$  then it is written as follows: $a<_f b$. All immediate greater elements are supposed to have unique labels. 

\textit{Tuple ordering principle.} In mathematics, partial order is represented as a binary relation  $\{ \langle a,b \rangle | a<b \}$. To represent a poset it is necessary to define another set. In COM, partial order is represented by tuples themselves with no need in any additional set. The connection between these two representations is established by the tuple ordering principle which postulates that a \textit{tuple is immediately less than any of its members}: 
\begin{equation} 
\mbox{[tuple ordering] }
\langle \ldots,f:e,\ldots\rangle <_f^1 e
\end{equation}

\noindent 
Equivalently, an element is immediately greater than any tuple where it is a member:  $e>_f^1 \langle \ldots,f:e,\ldots\rangle$. Labels correspond to dimensions and top element corresponds to empty tuple: $\top=\langle\rangle$. This principle can be used to represent an existing poset by writing each element as a tuple of its immediate greater elements. 

\textit{Function ordering principle.} The tuple ordering principle is written in terms of functions as follows: 
\begin{equation} 
\mbox{[function ordering] }
f(e) >_f^1 e
\end{equation}

\noindent 
It means that function output value is immediately greater than the function input value. Given a poset, it can be represented by functions corresponding to labels. Conversely, a number of functions can be represented as a labeled poset (if they satisfy this structural constraint).

\subsection{Inclusion}
\label{3.2}

\textit{Tuple inclusion principle.} A tree of sub-elements is represented by \textit{strict inclusion} relation  $\subset$. If  $a \subset b$  ($a$ is included in $b$) then $a$ is called a \textit{sub-element} and $b$ is called a \textit{super-element}. Sub-elements are said to be included in their super-elements. A connection between tuples and inclusion relation is established by the tuple inclusion principle: 
\begin{equation} 
\mbox{[tuple inclusion] }
\langle super:s | \ldots\rangle \subset^1  s
\end{equation}

\noindent 
It means that an extended element is immediately included in its base element. Since \textit{super} is a normal dimension, inclusion $\subset$ is a subset of partial order $<$: 
\[
a<_{super}^1 b \Leftrightarrow a \subset^1 b
\]

\noindent 
In other words, some instances of partial order which are labeled by \textit{super} are also instances of inclusion relation. Top element of the lattice  $\top=\langle \rangle$  is also referred to as a \textit{root} of the inclusion hierarchy. 

A \textit{nested set}  $\langle R,\subset \rangle$  is a set $R$ with an inclusion $\subset$ established on its elements. It is essentially a tree of elements where the root is the empty tuple and each element has many extensions but only one super-element. Given a nested set  $\langle R,\subset \rangle$, it can be represented as a number of tuples each defined as an extension of its immediate super-tuple: 
\[
e \subset^1 a \Rightarrow e=\langle super:a|b \rangle
\]

\textit{Type constraint.} Inclusion relation is not simply a subset of partial order. They are connected via \textit{type constraint} which is a structural analogue of the function extension principles (2): 
\begin{equation} 
\mbox{[type constraint] }
e \subset^1 a \land a <_f^1 c \land e <_f^1 d \Rightarrow d \subseteq c
\end{equation}

\noindent 
Its purpose is to exclude (semantically meaningless) situations by guaranteeing that \textit{all} constituents of a sub-element are included in the corresponding constituents of the super-element: 
\[
e \subset a \Rightarrow \forall f,e.f \subseteq a.f
\]

\begin{figure}
\begin{center}
\includegraphics[width=80mm]{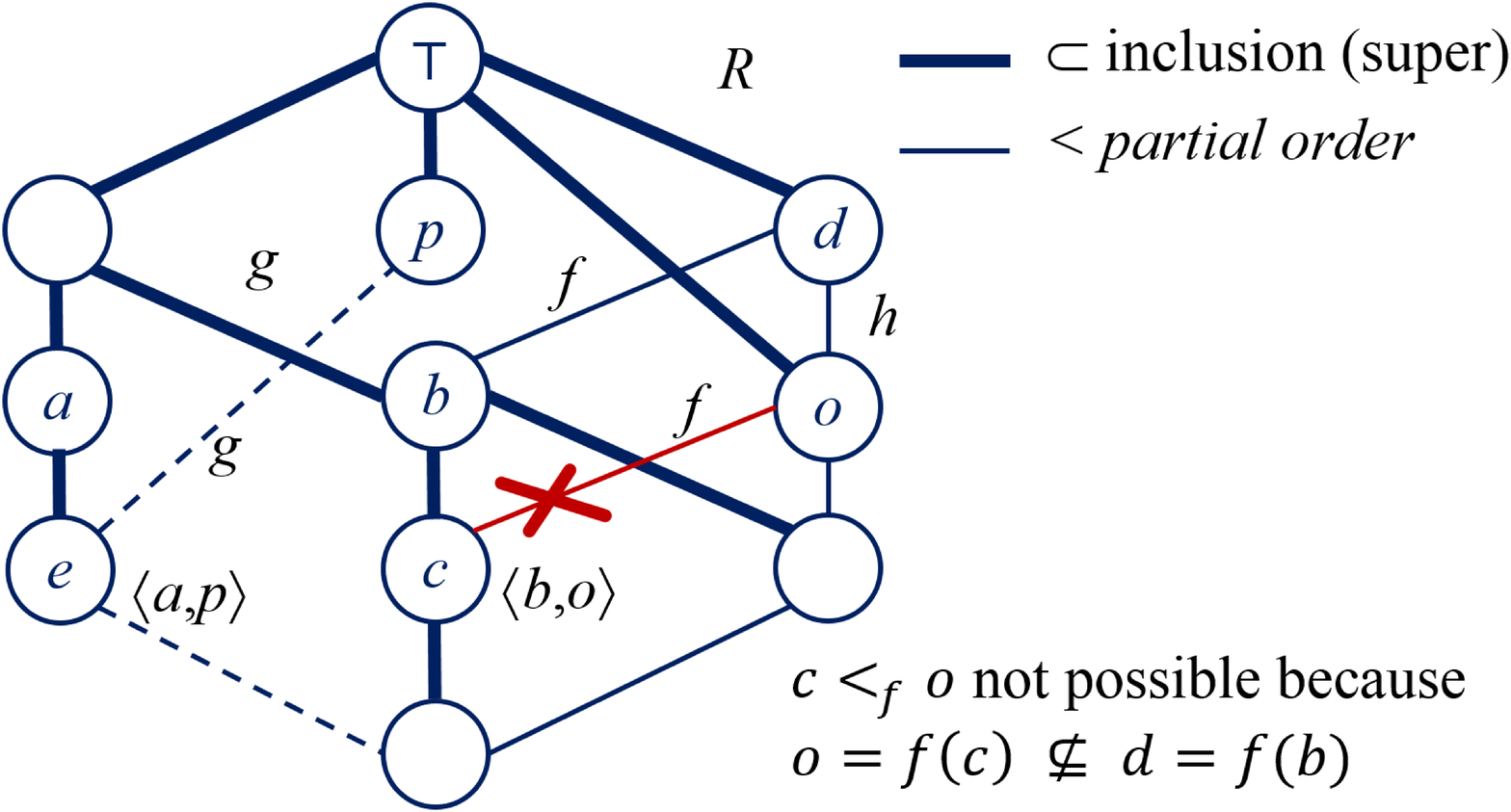}
\caption{Nested partially ordered set (Hasse diagram)} \label{figure02}
\end{center}
\end{figure}

For example (Fig.~\ref{figure02}), function $f$ returns $d$ for element $b$,  $f(b)=d$, and therefore all extensions of this function must return some extension of the element $d$. In particular, function $f$ must return some extension of $d$ for element $c$, and $f$ cannot return $o$ for $c$. This can be fixed if $o$ is made an extension of $d$ so that it conforms to the diagram in Fig.~\ref{figure01}. The use of the type constraint for overriding fields is demonstrated in Section 4.2. 

Type constraint can be described in terms of category theory and functors. A functor is a structure preserving mapping between two categories. A nested set can be interpreted as a category with elements as objects and instances of inclusion relation as arrows (morphisms). A (non-super) function is a (covariant) functor which maps this element to some other element. The goal of the type constraint is to preserve the inclusion structure, that is, output elements returned by the same function must have the same structure as input elements. In our case, input elements have nested structure and hence output elements also must have nested structure. 

Type constraint and function extension principles are similar to covariance and contravariance of parameters in programming languages. More specifically, it is analogous to \textit{covariant return types} where a more specific overriding method can return a more specific type than that returned by its base method. Also, many programming languages support this mechanism for generic parameters. For function types, covariance and contravariance were first described in \cite{Car84} where it was observed that it is safe to use a function that takes more general arguments and returns a more specific type than the overridden function. 

\subsection{Nested Partial Order}
\label{3.3}

\textit{Nested partially ordered set}  $\langle R,\subset,< \rangle$  is a set $R$ with strict partial order $<$ and strict inclusion $\subset$ relations established on its elements which satisfy type constraint (6). A database can be formally defined as a nested partially ordered set. More specifically, a \textit{concept-oriented database} is a finite set of elements $R$ satisfying the following conditions: 

\begin{itemize} 
\item any element is an identity tuple with an associated entity tuple according to the definition (1) 
\item any element is less than its tuple members (function outputs) so that $R$ is a partially ordered set according to the tuple ordering principle (3) 
\item any element is an extension of some super-element so that $R$ is a nested set according to the tuple inclusion principle (5) 
\item all elements satisfy the type constraint (6) 
\end{itemize}

This structure can be represented as a conventional finite lattice where each element has one upward path to the top element interpreted as set nesting (inclusion). An example of such structure is shown in Fig.~\ref{figure02} where inclusion tree (of super-dimensions) is drawn by bold lines. Also, there are two kinds of instances of partial order relation corresponding to identity tuples (solid lines) and entity tuples (dashed lines). Solid lines define the data itself, that is, what is stored and passed as values. Dashed lines define shared data, that is, what is stored persistently. An independent but very important mechanism is that this structure allows for labels with duplicate names but it restricts its use according to the type constraint (structural analogue of the function extension principle). Another way to visualize nested partially ordered sets is to show it is a nested Euler diagram with both sets and elements partially ordered or a tree representing inclusion relation with nodes partially ordered \cite{Sav14b}. 

Nested partial order is a structural approach to representing a concept-oriented database. It is an alternative to the functional approach described in the previous sections and syntactic approach based on the concept-oriented query language. An advantage of this representation is that it emphasizes structural aspects of the model (relationships among elements) but its disadvantage is that it is too abstract for implementation purposes where functional and syntactic approaches are mores suitable. But probably the most serious drawback of this representation is that it does not distinguish between data elements and sets, particularly, it does not say anything about differences between schema and instances. This issue is considered in the next section. 

\section{Sets}
\label{sets}

In the previous sections we described the structure of data elements but we did not say anything explicitly about the sets these elements are in (by assuming that all elements exist within one set). The goal of this section is to introduce a set as an explicit construct of the model.

\subsection{Structure of Sets}
\label{4.1}

Sets in COM provide a mechanism for declaring certain constraints on the structure of data elements. This means that if an element is a member of some set then it has to satisfy the constraints associated with this set and it cannot have arbitrary structure anymore. Since the structure is described by functions (an element is defined as a number of mappings to other elements), sets should restrict functions that can be used to describe these elements. 

\textit{Set} is defined very similar to data elements as one tuple of identity functions and an associated tuple of entity functions: 
\begin{equation} 
S=\langle f_1:F_1,\ldots,f_n:F_n \rangle ( g_1:G_1,\ldots,g_m:G_m )
\end{equation}

\noindent 
where  $f_i:S \rightarrow F_i$, $i=1\ldots n$,  and  $g_j:S \rightarrow G_j$, $j=1\ldots m$  are identity and entity functions, respectively. Note that sets are defined in terms of other sets treated as elements of the model. Essentially, a set is a combination of other sets which play a role of ranges for its functions. And there is only one way to create a new set: specify its functions. 

\textit{Structure of sets.} We can forget about set members and treat sets as normal elements. Then all properties of data elements and constraints applied on their structure are also valid for sets. In particular, we assume that all sets (treated as elements by ignoring their members) are partially ordered and hence we apply to them the corresponding terminology: greater and lesser sets, super- and subsets, primitive sets, top and bottom sets. The only difference is in the interpretation: instead of data elements we use sets and instead of function elements we use function declarations via their domains and ranges. 

\textit{Set types.} Set type involves two parts: a description of the mechanism for storing elements (like a table consisting of rows or a hash map), and the type of elements. In data modeling (in contrast to programming), only the latter is normally important. This means that a set type is defined via the type of its elements while the implementation of the storage is provided by the system. For example, we could say that a model needs a set of bank accounts without specifying what kind of storage this set will use. The storage for this set can be chosen separately like row-store or column-store. The type of data elements in COM is specified via concepts (Section 2.3) and also concepts are used to describe set types (by assuming some default storage type). In COQL, a set of elements of certain type is written as a concept name in parentheses. For example, \texttt{(Account)} denotes a set of bank accounts. 

\textit{Set extension.} Just like data elements, a set can extend another set if one of the identity functions represents a base set. In its limited form, this mechanism can be used to add new functions to an existing set by inheriting the base functions. However, it is more general than the classical extension mechanism because it supports object hierarchies and a more general mechanism for overriding functions.

\subsection{Database}
\label{4.2}

A \textit{concept-oriented database} is a number of set elements and data elements. Both of them have nested partial order relation as a constraint on their structure. Data elements of the database are represented by the functions of the set elements. Sets without function definitions (without data) define a schema. Thus defining a database schema is reduced to declaring new sets in terms of already existing set declarations. Syntactically, it is done by defining concepts. An empty database has only primitive sets. 

There is only one way to declare a new set which consists in specifying the following components: 

\begin{itemize} 
\item exactly one superset  $S$  (if not specified explicitly then by default it is assumed to be the root) 
\item zero or more range sets  $F_i$  for identity functions with unique names  $f_i$ 
\item zero or more range sets  $G_i$  for entity functions with unique names  $g_i$ 
\end{itemize} 

There are two kinds of primitive sets: sets that define values like numbers or strings, and sets that provide a mechanism of referencing. The sets of the former kind provide conventional data types but they do not have any entity functions. The sets of the latter kind appear in explicit form only in COM (implicitly they should exist in any system that provides persistence or representation by-reference). Essentially, such a set is a provider of generic references for its subsets and in a simplified form it could be declared as follows: 

\begin{alltt}
\textbf{CONCEPT} Reference 
  \textbf{IDENTITY} 
    Integer reference 
  \textbf{ENTITY} 
    Byte getData(Integer offset)
\end{alltt}

A data management system can provide several types of such primitive references for different purposes. For example, there could be different references for small sets and big sets, for local storage and network distributed storage and so on. What is important is that for every new set, a superset has to be chosen and this superset defines whether the new set will be a conventional data type (passed by-value) or a reference type (passed by-reference). More complex system can provide a possibility to develop concepts which implement some user-defined logic of persistence and access (see \cite{Sav12c} for more information). 

\begin{figure}
\begin{center}
\includegraphics[width=80mm]{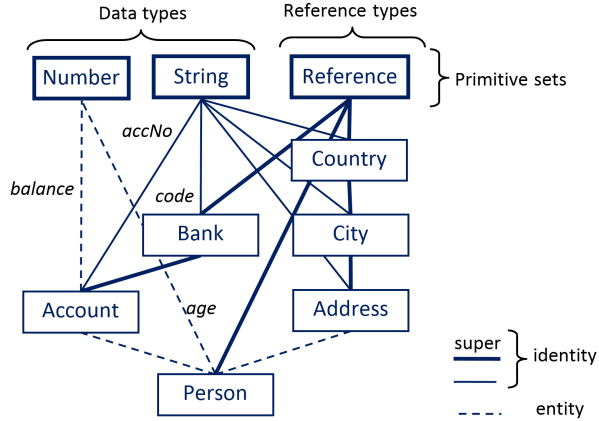}
\caption{Concept-oriented schema} \label{figure03}
\end{center}
\end{figure}

An example of a concept-oriented schema is shown in Fig.~\ref{figure03}. Identity functions are represented by solid lines: bold lines show super-dimensions and normal solid lines represent other identity dimensions. Any set that is intended to store data by-reference must have reference as a direct or indirect root. If this set has to have many instances for one super-element then it is necessary to add other identity dimensions. For example, the set \texttt{City} is included in \texttt{Country} but it also has an additional identity dimension so that many city instances can be created within one country instance. Note that the \texttt{Person} set does not have own identity dimensions and hence its instances will be identified by system references. Entity functions are shown by dashed lines and they return a value associated with the current identity but shared among all other elements. For example, every account has a balance and every person has an associated bank account and address. 

\subsection{Operations}
\label{4.3}

\textit{Product.} All possible elements of a set are defined via the product of all its identity sets (including the superset): 
\begin{align} 
\begin{split}
P &= S\times F_1\times\ldots\times F_n \\
&= \{ p=\langle s,f_1,\ldots,f_n \rangle | \forall i,f_i \in F_i,s \in S \}
\end{split}
\end{align}

\noindent 
Here  $S$  is a superset and  $F_1,\ldots,F_n$  are other greater identity sets. This definition of the product operation has the following properties: 

\begin{itemize} 
\item The arity of the product is equal to the number of identity sets (including the superset) what conforms to the conventional mathematical definition of product. For comparison, the arity in RM is equal to the sum of arities of the source sets. 
\item Entities of the source sets are not taken into account. 
\item Product operation defines only identities. Entity functions of the new set can be defined either explicitly for all element or defined as an expression in terms of other functions (derived dimension). 
\item Product is a lesser set with respect to all its source sets. Thus product is not an isolated set -- it exists within the structure of its source sets and is connected with them via functions. 
\end{itemize}

\textit{Subset.} Creating a subset is one of the most wide spread operations where it is necessary to produce a new set containing elements from one source set satisfying certain properties. A subset in COM is defined as a set without own identity but inheriting identity from the superset. Its elements are selected by providing an additional predicate  $u$  which has to be true for all selected elements: 
\begin{equation} 
E = \{ e=\langle super:s \rangle () | s \in S, u(s)=true \} \subseteq S
\end{equation}

\noindent 
Here $E$ is a subset of $S$ and each element from $E$ references some element from $S$ via its super-function without adding new functions (neither identity nor entity). By adding functions to the entity part we can define extended sets precisely as they are treated in object-orientation. Note again that a subset is not defined in isolation by abstractly selecting necessary elements: a subset is an element of the model with a certain position in its structure. 

\textit{Derived functions} are defined via expressions computing their output from the values returned by other functions in the database. Any derived function is defined in the context of the current element, accessed via a special \textit{this} function, and returns a value from its declared range. The simplest (row-based) derived functions use arithmetic expressions and dimensions of the current set. For example, we could define a function which returns the ratio between employee age and salary: 

\begin{alltt}
age2salary() = this.age() / this.salary()
\end{alltt}

\noindent 
Dot notation can be used to access functions of greater elements. For example, a function returning complete address simply concatenates several strings: 

\begin{alltt}
address() = 
  address.country.name() + 
  address.city.name()
\end{alltt}

\noindent 
However a really powerful mechanism for accessing arbitrary functions in the database is provided by a novel arrow notation described below in this section. 

\textit{Set operations.} These operations manipulate sets of elements rather than individual items. Two most important operations for set manipulations are project and de-project. \textit{Project}, denoted by right arrow, is applied to a set of elements and returns all distinct outputs of the specified function: 
\begin{equation} 
E \rightarrow f \rightarrow F = \{ f(e) \in F | e \in E \}
\end{equation}

\noindent 
In terms of partial order it is defined as follows: 
\[
E \rightarrow f \rightarrow F = \{ a \in F | \exists e \in E, e<_f a \}
\]

\noindent 
Since projection is a set, it contains any element only one time without duplicates. 

\textit{De-project,} denoted by left arrow, is an opposite operation which returns all inputs which are mapped to the elements from the argument via the specified function: 
\begin{equation} 
F \leftarrow f \leftarrow E = \{ a \in E | \exists e \in F, f(a)=e \}
\end{equation}

\noindent 
In terms of partial order it is defined as follows: 
\[
F \leftarrow f \leftarrow E = \{ a \in E | \exists e \in F, a<_f e \}
\]

\noindent 
De-project operation can be also defined in terms of inverted functions in category theory. If  $f(x):E \rightarrow F$  is a function then inverted function  $f^{-1}(x):F \rightarrow P^E$, where  $P^E$  is a powerset of  $E$,  returns a set of inputs for a certain output: 
\begin{equation} 
\mbox{[inverted function] }
f^{-1} (x)=\{ a \in E | f(a)=x \}
\end{equation}

\noindent 
De-project operation is then defined as a union of all outputs of the inverted function: 
\[
E \leftarrow f \leftarrow F = \bigcup_{e \in E}{f^{-1}(e)}
\]

\textit{Arrow notation.} Project and de-project operations can be applied to the result returned by the previous operation. This approach is referred to as \textit{arrow notation} because it is very similar to the conventional dot notation. The main difference and advantage of arrow notation is that it is intended for manipulating and navigating through sets rather than through instances in a graph. Also, this notation explicitly uses two directions for navigation via inversion (de-project). Another advantage is that it does not use joins and group-bys but rather relies on references which is very simple and natural approach. Dot notation is especially useful if we take into account semantics of references and the data structure in COM so that arrows are not simply navigation operations but rather operations having a significant semantic load like changing the level of details or change the generality level. 

\section{Semantics}
\label{semantics}

Assume that we have such basic mathematical constructs like tuple or set as well as more complex structures like poset or topology. Is it already a data model? No. In order to use a structure for data modeling it is necessary to define its meaning in common terms accepted in real life. This is done by defining data semantics and it is therefore of primary importance for data modeling. Earlier we have already described some basic semantics. For example, we postulated that datum (data element) is a value and values are represented by tuples. Below we will describe more complex semantic constructs. 

\textit{References.} COM assumes that a reference \textit{is} a value and in this sense we cannot distinguish between normal values like number 5 and references (so number 5 may well be a reference). What really turns a value into a reference is the possibility to provide access to other values. This possibility is integrated into any data element in COM via the duality principle. References are implemented via functions the implementation of which is hidden so we actually do not know where data is stored. 

\textit{Characterization.} Characterization is perhaps one of the most wide spread approaches to describe things where an element is supposed to have attributes which may take values. It exists in almost all other data models but in COM it is especially simple because it does not distinguishes between attributes storing primitive values, complex values, objects or multiple-valued attributes. There is only one basic way to directly characterize things in COM: to store (identities of) other things in the definition. Note also that any element can be characterized by other elements and, vice versa, it can characterize other elements (if it does not break structural constraints). 

\textit{Set-valued attributes.} It is a very important feature because many things are characterized by a subset of other things rather than by a single value. Many existing models  provide a straightforward solution to this problem by simply marking a field as a set-valued or collection. However, this approach is difficult to formalize and it leads to numerous problems in complex models with complex relationships. For example, what if the subsets of values used to characterize other elements are themselves characterized by set-valued attributes? The desire to formalize the mechanism of set-valued attributes was one of the main motivations for introducing the nested relation model (NRM) \cite{Sch86}. COM does not provide set-valued attributes as a basic feature because it makes the model more complicated without necessity. Instead, this mechanism is provided at higher level using the assumption that \textit{lesser elements represent subsets characterizing a greater element}. If it is necessary to have this characteristic as an attribute then it can be defined as a derived function which is equivalent to normal dimensions but returns a (computed) subset of elements. For example (Fig.~\ref{figure03}), all account holders of a bank can be returned by the following derived function of the \texttt{Bank} concept which deprojects this bank to the \texttt{Person} concept: 

\begin{alltt}
(Person) Bank::AccountHolders() \{ 
  return this 
    <- bank <- (Account) 
    <- account <- (Person) 
\}
\end{alltt}

\textit{Relationships.} A relationship is a dedicated data modeling constructs which is used to establish an association between data elements. The idea of using relationships was proposed in the entity-relationship model (ERM) \cite{Che76}. Although relationships provide very powerful means for conceptual modeling, their use has a number of drawbacks: (i) there exist many possibilities to relate the same elements and therefore the use of relationships is normally quite ambiguous and depends on the application \cite{Dah05}, (ii) it is not clear when to use relationships and when to use attributes, (iii) it is not clear what is the difference between relationships and normal data. For a unified model these are quite serious issues. On the other hand, relationships are known to be very useful and cannot be ignored. Therefore, COM does not provide relationships as a dedicated construct but rather uses the following principle: \textit{a common lesser concept is treated as a relationship or dependency connecting its greater concepts}. So if we know that some existing concepts are somehow related or depend on each other then we simply introduce a common lesser concept which connects them. For example, \texttt{Person} in Fig.~\ref{figure03} is a normal concept but in COM it is also treated as a relationship between people and addresses which are its greater concepts. 

\textit{Containment.} A specific feature of COM is that it provides two mechanisms for modeling containment (IS-IN) relation: by-reference using partial order and by-value using inclusion relation. In order to include an element in another element it is enough to reference it. This relation can be changed by changing this reference. A more specific type of containment is provided by inclusion relation. It is used for modeling identification hierarchies like postal addresses. Note that an element can be also viewed as consisting of its greater elements. However, this relation has completely different semantics and describes PART-OF relation: greater elements are parts within their common lesser element. 

\textit{General-specific and inheritance.} COM does not provide this relation as a separate construct. Instead, COM assumes that it is a particular case of containment: \textit{to be contained within some element means to be more specific element and to inherit properties of the container}. Thus all lesser elements and sub-elements are interpreted as more specific elements with respect to their greater elements and super-elements. The classical inheritance is modeled by creating a new concept without own identity and adding only some entity dimensions. In the general case, however, an element may have many more specific extensions inheriting its properties. 

\textit{Multidimensional space.} Another important semantic interpretation consists in thinking of elements as existing in a multidimensional space. This approach is used in multidimensional models where the roles of axes and coordinates are assigned to the elements of the model. COM supports this interpretation by assuming that \textit{lesser elements are points while greater elements are coordinates}. For example, if a book element references a publisher then the book is interpreted as a point while the publisher is one of its coordinates. Here again we see that an element can be simultaneously a point with respect to its greater elements and a coordinate for its lesser elements. COM is more flexible than the standard multidimensional models because it does not rely on predefined cubes, dimensions, facts and measures which are application-specific roles. Data is thought of as originally existing in a multidimensional space so that it is always possible to say what coordinates this element has and how many dimensions this schema has. The partially ordered schema in COM can also be viewed as a generalization of star, snowflake and snowstorm schemas where lesser concepts correspond to fact tables and greater concepts describe detail tables. 

\textit{Graphs.} A graph can be viewed either as a formal structure (like poset) or as a semantic model where things are supposed to be connected by relationships. Both uses are widely used in different models. Yet, not in COM. Although references in COM are viewed as a basic mechanism for connecting elements and therefore can be interpreted as edges in a graph, it is a wrong interpretation. COM assumes that if two elements are connected via a reference then they by definition belong to different levels and play different roles: container-member, coordinate-point, master-detail or general-specific. However, a graph metaphor is known to be very popular and useful in data modeling so the question is how two elements can be connected by an edge? It is done using the following principle: \textit{common lesser elements in COM are interpreted as edges connecting their greater elements interpreted as vertices}. 

\section{Discussion and Related Work}
\label{discussion}

\textit{How identities are different primary keys and oids?} Identity is by definition a value. In contrast, a key is a subset of entity attributes and hence is not a value as such. Identities are much closer to conventional (primitive) references and oids \cite{Sav11a}. The difference is that identities (like primary keys) have arbitrary domain-specific structure. 

\textit{Should identities be modeled?} Essentially the question is whether identities are part of the problem domain and whether they are data at all. These questions are quite important because many models (mostly object-oriented) assume that entity identifiers have a primitive form and have to be provided by the platform and hence identities are not part of the problem domain. Also some models assume that only primitive values should be used while all more complex structures should be modeled by objects. COM assumes that identities are data just because they are values. Since values play primary role in data modeling there should be dedicated means for their modeling. The question whether some kind of data belongs to the problem domain or to the system is about the level of abstraction. Consequently, both identities and entities can belong to the problem domain or be provided by the platform. 

\textit{Is entity a value?} No, entity is not a value. Only constituents of an entity returned by individual entity functions are values. Writing an entity as a tuple is a convenient way to represent related elements but in contrast to identity tuples, entity tuples cannot be passed or stored as one whole (one value). Entity constituents are values which are accessible using one identity but there is nothing else that unites them. In particular, they are not supposed to be stored together side-by-side as is implicitly assumed in many other models. 

\textit{Why concepts?} Concepts generalize classes and provide a number of benefits. They allow for modeling simultaneously values and objects by hiding this separation behind one name. It is important if the structure changes because element (fields, variables, parameters etc.) types do not depend on the internal structure of concepts and distribution of functions between identity and entity parts. Concepts also significantly simplify data modeling at conceptual level because now there is no need to distinguish between value types and object/entity types. It is enough to have only one construct for all data typing tasks. 

\textit{Why functions?} Functions provide a convenient mathematical abstraction for the underlying data access and storage mechanisms. In other words, we do not actually know where the data is stored and how it is being accessed but we know that for each input value we can get the corresponding output value by using function names. The use of functions is very natural in the context of column-stores providing higher performance for analytical workloads because functions can be very naturally mapped to the column-oriented representation of data. Note that functions are storage elements, that is, all data is stored in functions as opposed to storing data in rows of sets. 

\textit{COM vs. logical models.} Due to their generality, both the functional \cite{Bun97} and the logical views on data can be treated as the lowest common denominator for other data models. This means that differences between them are of fundamental character and it is hardly possible to justify that one of them is better than the other. We would like to describe only one such fundamental feature distinguishing COM (which belongs to the class of functional data models) and models based on the first-order logic: how they represent connectivity between things. 

\begin{figure}
\begin{center}
\includegraphics[width=80mm]{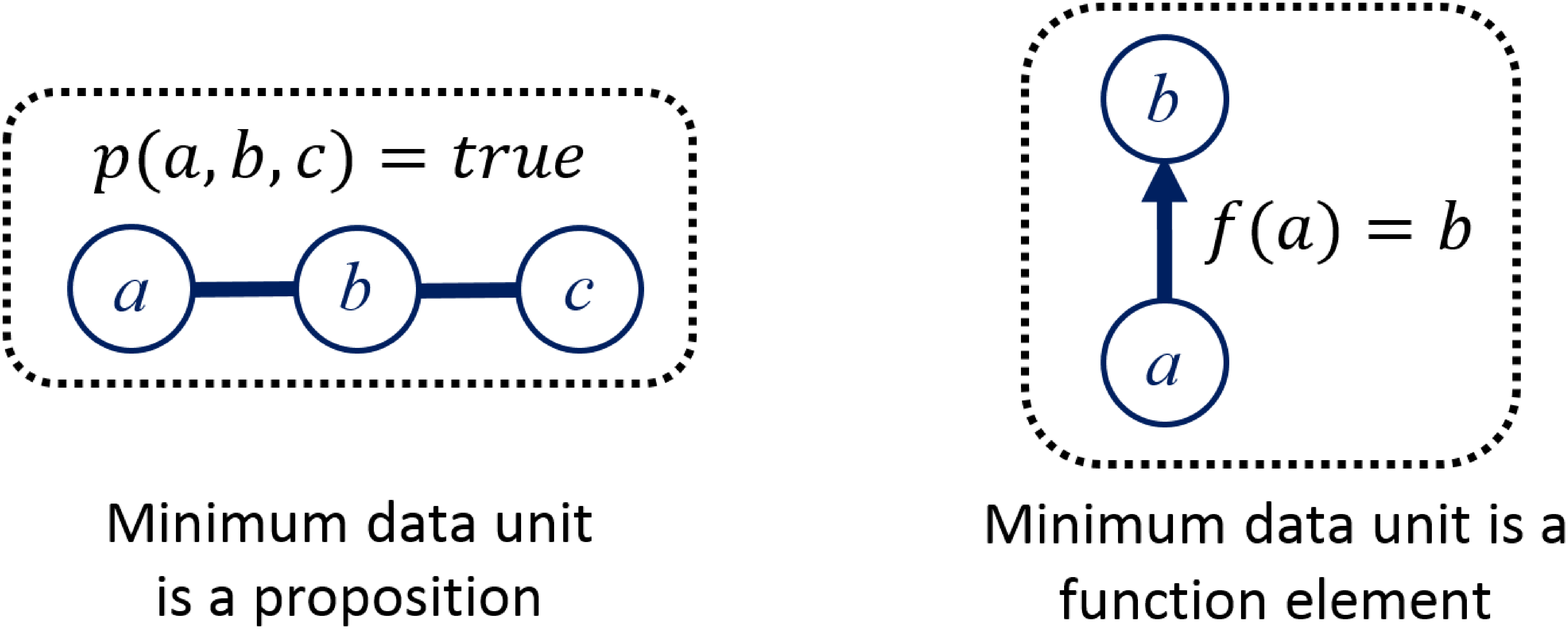}
\caption{Connectivity in functional and logical models} \label{figure04}
\end{center}
\end{figure}

In logical data models, if we want to say that things are somehow connected then we define a predicate which is true for these things (Fig.~\ref{figure04} left). For example, if we want to describe which accounts belong to which banks then it is done via a two-place predicate \texttt{belongsTo(acc, bank)}. The most important consequence is that all such relationships are symmetric (horizontal) and the predicates may have any arity including unary, binary and $n$-ary predicates. Another property is that this model has only two levels: atoms and propositions (predicates). In particular, it is not possible to make propositions about other propositions. Of course, this can be overcome by introducing additional roles and treatments for predicates, for example, a unary predicate could represent set membership relation (one predicate then represents one set) and binary predicates could represent symmetric relationships like roles in description logics (DL). Yet, the usefulness of these extensions only emphasizes that having only atoms and predicates is not enough for a good data model. 

The main construct for describing connectivity in functional data models is that of function. If two things are related than one of them is mapped to the other using some function (Fig.~\ref{figure04} right). For example, an account element is mapped to the bank element using the function \texttt{belongs}. Importantly, it is not equivalent to having binary predicates. It is possible to model functions by means of predicates but this should viewed as a workaround in the case functions are not directly supported. One difference between functions is predicates is that the former are directed (function is a mapping) while the latter are symmetric. One immediate advantage of this property is that functions can be composed and hierarchies have natural support while the composition of predicates is performed quite differently. Note also that functions can well support the horizontal symmetric connectivity as it is done by predicates. If an element is mapped to many other elements using different functions then it is treated as a tuple of other elements which is analogous to a proposition about these elements. In this sense, the functional paradigm (and COM) can cover both dimensions: vertical (a constituent might have its own constituents), and horizontal (an element can be composed of many constituents). In formal logic, they are modeled using predicates. In the functional approach (and in COM), they are modeled using two dedicated constructs: functions and sets (of tuples). 

\textit{COM vs. the functional data model (FDM).} The goal of FDM \cite{Sib77} is to provide simple but general means for manipulating data by relying on the mathematical notion of function. COM also uses this notion and hence the question is what the differences between these two models are. First, FDM uses classical approach to the treatment of data elements and data typing while COM distinguishes between identity functions and entity functions. Second, FDM functions are mappings from only primitive entity identifiers to either values or other entity identifiers. In particular, arbitrary values cannot be used as inputs to FDM functions. COM functions are mappings between \textit{arbitrary} values which also can be references. Third, the structure of FDM functions is not constrained and it is an arbitrary graph. COM functions are constrained by partial order which is important for representing data semantics. In other words, a function in COM is not simply a mapping -- it is a semantic construct with many interpretations (output is more general than input, output is a coordinate for input, output is a characteristic of input, output is a container for input). Fourth, COM provides special super-functions and the principle of function extension which underlie several important mechanisms like object hierarchies, incremental overriding, and reverse overriding. Hierarchical organization of entities and property inheritance is provided in the Extended Functional Model (EFDM) \cite{Kul86} but it is a standard (object-oriented) approach to inheritance: an instance of a type is also an instance of its super-types, and a function which applies to a type also applies all of its subtypes. COM generalizes it by using object hierarchies and a more complex overriding strategy. 

\textit{Why only single-valued functions?} Functions in COM are not simply mappings between sets -- they have quite significant semantic load, that is, to define a function means to define meaning for a connection between two sets. Multi-valued functions make natural semantic interpretation more difficult. For example, if we want to interpret a function as returning a coordinate for this object along one axis then it is quite natural to allow for only one output because an object in real world cannot simultaneously have several locations. Also, if an element has several coordinates or characteristics then it is difficult to answer the question about dimensionality. Indeed, if an element has  $n$  single-valued properties then it is naturally positioned in an $n$-dimensional space. But if one of these properties can return 2 or more outputs then the dimensionality obviously grows because they can be varied independently. Hence, the object is positioned in a space with variable dimensionality where dimensions are not clearly identified which is not a good feature. Similar difficulties appear in any model that wants to introduce multi-valued, relation-valued properties or nested relations and this mechanism leads to numerous complications. For that reason, COM does not use multi-valued functions at the level of schema, that is, a function always returns a single value. Yet, in practice, multi-valued characteristics and relationships are met very frequently so we cannot simply ignore them. To model them COM uses the principle that \textit{if a thing is characterized by several other things then these characteristics are lesser elements}. Essentially, this means that a multi-valued function is a reversed function and hence they can be defined via de-projection operation. In addition, COM uses derived functions which are defined via other functions and are allowed to have multi-valued output. Therefore, multi-valued properties can always be defined via derived functions which return either lesser elements or sets of elements retrieved from other sets in the schema using arbitrary query. 

\textit{COM vs. the relational model (RM).} COM can be reduced to RM by assuming that all sets have only primitive greater sets. Therefore, most formal aspects of the relational algebra will also work in COM. If greater sets are sets of values with arbitrary structure then we get the object-relational model (ORM) \cite{Kim90}. In both cases we get a kind of two-level schema with domains and relations at two levels. Interpreting COM identities in relational terms is ambiguous. One way is to assume that primary keys correspond to COM identities. However, primary keys are composed of normal attributes which have the same status as all other attributes while in COM identities are treated quite differently. Another way is to assume that COM identities correspond to surrogates which in addition may have user-defined structure modeled by identity classes. This analogy is much closer to the purpose of COM identities however surrogates and row identifiers are not part of RM in its classical formulation and the need in this mechanism is a quite controversial issue. In this context, COM provides additional arguments in favor of having primitive identifiers in a good data model. Another significant difference of COM from RM is that COM is inherently semantic model which provides various general semantic interpretations to its constructs rather than being a formal algebra for manipulating tuples and relations. 

\begin{figure}
\begin{center}
\includegraphics[width=80mm]{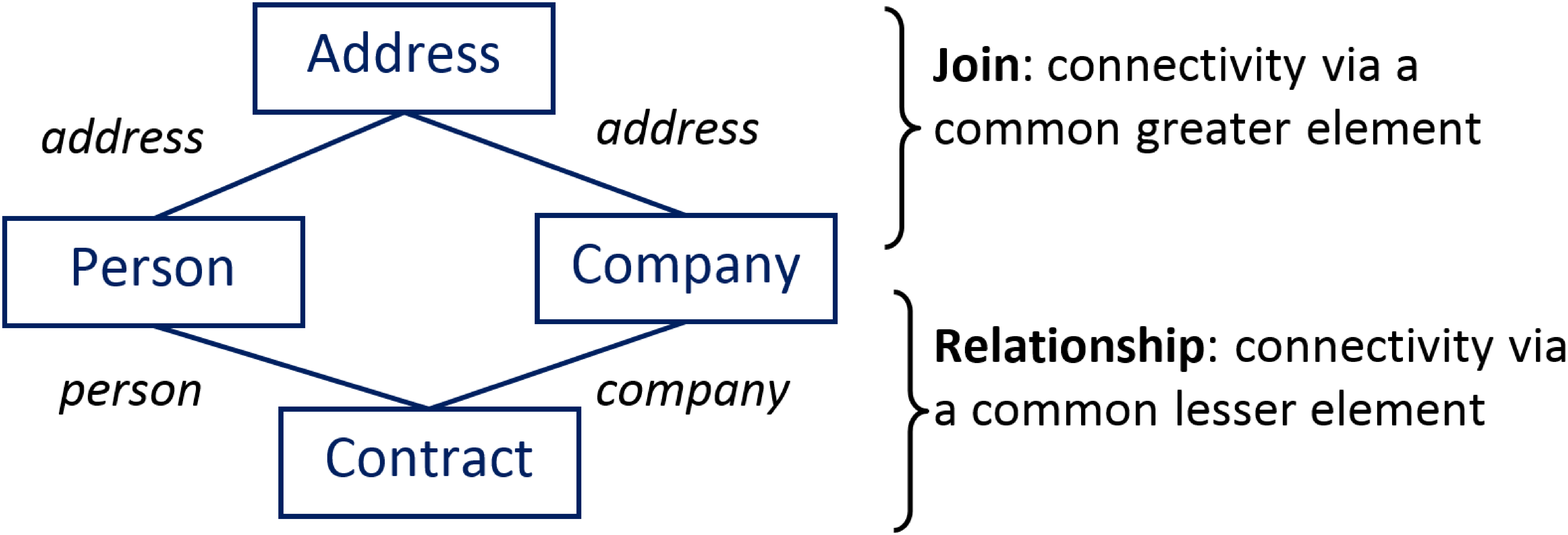}
\caption{Joins vs. relationships} \label{figure05}
\end{center}
\end{figure}

\textit{Where are joins?} Joins can be used in COM by imposing constraints on product operation. For example (Fig.~\ref{figure05}), we could find the product of sets \texttt{Person} and \texttt{Address}, and then select only elements that have equal values in some fields: 

\begin{alltt}
(Person p, Address a | p.address == a)
\end{alltt}

In most cases however, joins are used to retrieve elements connected via references but if the mechanism of referencing is not inherently supported (as it is in RM) then the only way is to implement it manually at the level of queries by means of joins. The use of joins for implementing references has significant drawbacks \cite{Sav16b,Sav12a}: (i) it is a low level and error-prone operation \cite{Atz13}, (ii) it lacks semantics by easily producing wrong results because the intention of the operation is not obvious from its syntax, (iii) it is a cross-cutting concern because the same fragments are explicitly written in many queries, (iv) joins describe the logic of connectivity which appears at the same level as the domain-specific logic of retrieval, (v) joins are not analytics friendly. Therefore, COM does not encourage the use of product operation with filters for that purpose. Instead, arrow notation should be used which inherently supports references and set operations. 

There are more complex patterns of connectivity and COM distinguishes two of them (Fig.~\ref{figure05}): \textit{common greater values} and \textit{common lesser values}. The former pattern means that two data elements or concepts are related if they have a common greater element. The latter pattern means that two elements are related if they have a common lesser element. Obviously, the first pattern corresponds to joins and the second pattern corresponds to relationships (discussed in Section 5). Thus COM establishes a nice duality between two connectivity patterns. These two patterns provide higher level mechanisms of connectivity based on references. For example, \texttt{Person} and \texttt{Company} in Fig.~\ref{figure05} are not directly connected in the schema. Yet, they can be connected indirectly and there are only two general ways: to find a common greater element like a common \texttt{Address}, or to find a common lesser element like a common \texttt{Contract}. Both patterns can be used in queries for finding related elements without classical joins: 

\begin{alltt}
(Person) -> (Address) <- (Company) 
(Person) <- (Contract) -> (Company)
\end{alltt}

This duality of joins and relationships is quite important for understanding the nature of connectivity. In particular, the connectivity via lesser elements (relationships or dependencies) provides a basis for the mechanism of inference in multidimensional space \cite{Sav12b,Sav11b}. It is an important mechanism because it allows for going beyond numeric analysis and doing in multidimensional space what has always been a prerogative of logic-based models. 

\textit{Universal relation model (URM)} \cite{Ley89}. The goal of this extension of RM was to relieve the user of the need for specifying concrete join conditions and to achieve access path independence where a query is written in terms of only attributes. For example, a query for getting all wheel suppliers is written as follows: 

\begin{alltt}
retrieve (Suppliers) 
where Parts = 'wheel'
\end{alltt}

\noindent
Note that this query does not provide any indication how suppliers are connected with parts -- it is the task of the system to translate it into the logical structure of relations. The main idea of URM consists in regarding the database as a single relation, called the universal relation, so that all other relations are its projections. Essentially, the universal relation is a kind of canonical representation where a database is viewed as one whole rather than as a flat set of relations. Such a holistic view is absent in RM but is important for many data management tasks: easy querying, inference, schema matching, analysis, consistency and others. Yet, the assumption of universal relation was shown to be incompatible with many aspects of the relational model and it did not result in a successful foundation for data modeling. 

\begin{figure}
\begin{center}
\includegraphics[width=80mm]{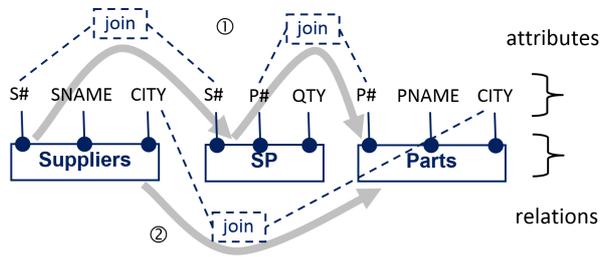}
\caption{Constraint propagation in URM} \label{figure06}
\end{center}
\end{figure}

\begin{figure}
\begin{center}
\includegraphics[width=80mm]{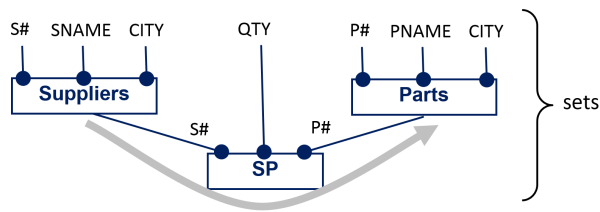}
\caption{Constraint propagation in COM} \label{figure07}
\end{center}
\end{figure}

The problems raised in the context of URM are still very actual and at the general level COM provides an alternative solution to them by relying on order-theoretic basis (as opposed to relational algebra). More specifically, bottom concept in COM can be viewed as an analogue of the universal relation. The difference is that the universal relation is defined over a set of (primitive) attributes while COM schema has many levels within a partially ordered set. This difference is illustrated in Fig.~\ref{figure06}. Assume that there are three sets of \texttt{Suppliers}, \texttt{Parts} and \texttt{SP} which is a relationship between \texttt{Suppliers} and \texttt{Parts}. The task is to automatically retrieve all parts delivered by some suppliers. URM analyzes joins between relations and tries to propagate initial constraints from suppliers to parts. In this example, there exist two paths. The first one uses a join between \texttt{Suppliers} and \texttt{SP} on \texttt{S\#} and then a join between \texttt{SP} and \texttt{Parts} on \texttt{P\#}. The second path uses a join between \texttt{Suppliers} and \texttt{Parts} on \texttt{CITY}. The problem is that such a propagation through joins is ambiguous: joins are undirected and there are many possible paths from the source to destination. COM relies on the structure of references by partially ordering the relations and propagation through lesser elements. In this example (Fig.~\ref{figure07}), the system builds the propagation path from \texttt{Suppliers} down to \texttt{SP} and then up to \texttt{Parts}. Such paths have clearer semantics and less ambiguity \cite{Sav11b,Sav12b}. 

\textit{Why partial order?} The only known work where partial order relation is laid at the foundation of data modeling is \cite{Ray96} where ``partial order database is simply a partial order''. However, this approach relies more on formal logic and it is focused on manipulating many partial orders. One reason why partial order is not widely used in data modeling is that it is a quite strong constraint for possible structures. Indeed, self-references and cycles are very common in database designs especially in the presence of graph databases. Therefore, a natural question is whether it is a good idea to use partial order and why it was introduced at all? Shortly, partial order in COM is introduced for the same reason trees are used for inheritance. More specifically, references in COM have much higher semantic load than in other models: they represent more general elements, containers, and coordinates. If we want to have a semantic model (and COM is intended to be a semantic model) then we need to exclude meaningless situations where an element is contained in itself, an element is a coordinate for itself and so on. And it is precisely where partial order works perfectly. If for some reason it is necessary to have a self-reference or a cycle then it always can be done by scarifying semantic consistency. All queries will still work except for smarter mechanisms like inference which can be confused by cycles as well as applications and processes which assume partial order like OLAP analysis. A better solution consists in marking dimensions which are known to reference a more specific element (while all dimensions by default reference more general elements). Such markers essentially resolves semantic cycles in element definitions by retaining the model in consistent state. 

\textit{COM vs. conceptual modeling.} COM does not completely belong to the class of traditional conceptual models such as Entity-Relationship Model (ER), Universal Modeling Language (UML) or Object-Role Modeling (ORM) \cite{Hal10}. These models have higher level of description and wider scope of applicability by allowing to represent more domain knowledge using richer sets of data modeling constructs. In contrast, COM is positioned as a logical data model with strong support for conceptual modeling. It is important that COM not only brings more semantics to the logical level but also generalizes some semantic relations by simplifying the process of modeling as described in Section 5. Thus the main advantage is that conceptual modeling is not a separate layer with its own modeling constructs but rather is an integral part of the logical level of the model. 

\section{Concept-Oriented Model for Data Wrangling}
\label{datacommandr}

In this section, we describe one possible application of COM which has been implemented in a framework for agile data transformations and manipulations, called DataCommandr \cite{Sav16a}. It is a data processing engine behind ConceptMix \cite{Sav14a} -- a tool for self-service data transformations. Data manipulations in DataCommandr are described using the Concept-Oriented Expression Language (COEL). The aim of COEL is to provide a simple and natural language for describing complex queries against multiple tables without using such classical set-oriented operations as joins, group-by and aggregation. The closest analogue of COEL is Data Analysis Expressions (DAX) used in Microsoft Tabular Model \cite{Rus12}. Our main research goal when implementing DataCommandr and COEL was to demonstrate that it is possible and even easier to work with such a functional (column-oriented) approach without the need in having set operations like joins and group-by. 

Let us illustrate how this new approach differs from the classical view on data manipulations. Assume that we have already some data stored in multiple tables. Our goal is to derive new data in the form of one or more tables each having some columns. The currently dominating approach is to define new tables in terms of already existing (or previously defined) tables, that is, a new table is a function of other tables: $T=f(T_1,\ldots,T_n)$. Normally, this definition is made at the level of one row, that is, an output row is defined as a function of input rows: $r=f(r_1,\ldots,r_n)$. The looping strategy is then implemented automatically by the underlying system like RDBMS, ETL or MapReduce. Note that the main unit of definition and the return value of expressions in such languages is one row (tuple). In contrast, the main unit of definition in COEL is that of column, that is, a column is defined as a function of other columns: $C=f(C_1,\ldots,C_n)$. Just as new tables are normally defined at the level of rows (tuples), new columns in COEL are defined at the level of values (in both cases the goal is to exclude explicit loops from the language). In other words, an output value in COEL expressions is defined as a function of input values: $v=f(v_1,\ldots,v_n)$. Thus instead of using set operations and returning tuples, COEL expressions use column operations and return values. In this sense, COEL is similar to how formulas are defined in typical spreadsheet applications with the main difference that it defines columns via other columns rather than cells via other cells. 

\begin{figure}
\begin{center}
\includegraphics[width=80mm]{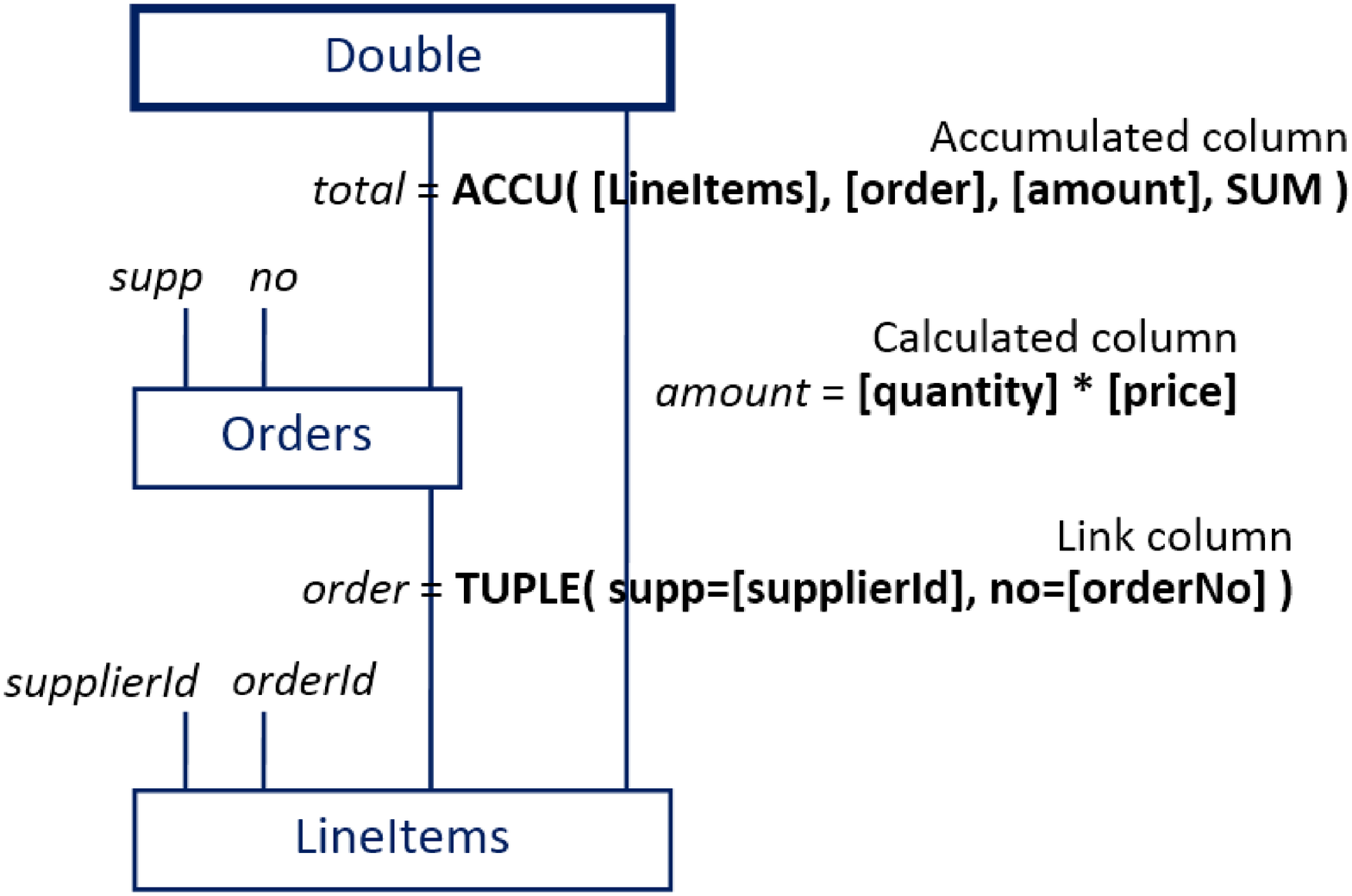}
\caption{Concept-oriented expression language} \label{figure08}
\end{center}
\end{figure}

A table in DataCommandr is created as a Java object without data and without any essential properties because all the data is stored in columns. COEL is used in tables only for representing constraints (analogous to SQL WHERE clause). However, table constraints are actually represented as functions returning true or false and therefore have the same form as COEL column definitions described below. If some columns already store data (for example, loaded from a file) then we can easily define a new \textit{calculated column} as a function which returns one primitive output value for each input row. For example (Fig.~\ref{figure08}), we could compute a column with the total amount for each line item in a purchase order by defining the following column:

\begin{alltt}
DcColumn amount = \\ createColumn("amount", "LineItems", "Double"); 
amount.definition = \textbf{"[price] * [quantity]"}; 
\end{alltt}

\noindent
The first line creates a Java column object with the name \texttt{amount}, input table \texttt{LineItems} and output (primitive) table \texttt{Double}. The second statement provides a definition for this new column as a COEL expression (written in bold). Note that COEL expressions refer to other elements by name in square brackets. Of course, this column could be easily defined in SQL by adding this formula to the SELECT clause. Yet, from the conceptual point of view, it is important that the \texttt{amount} column is thought of as an independent data modeling unit with its own definition. In other words, we have made an important conceptual shift by switching from ``table views'' to ``column views''. It is easy to add columns, delete columns, index columns and perform column-oriented operations. A table in this approach does not have its own data at all -- it is an element of the schema level rather than the data level. 

Defining new derived columns within one table is easy. What is not trivial is working with multiple tables. For example, assume that we have two tables with purchase orders and line items loaded from two CSV files. These tables have only primitive columns and therefore not directly connected but we would like to define a new column by accessing data from the both tables. Normally the only possibility is to apply join operation and produce a new table with the data from the two source tables. DataCommandr provides a different mechanism by defining \textit{link columns} which explicitly connect two tables without producing a new table. To define a link column it is necessary to return a tuple rather than a primitive value. For example, if we want to define a column in the table \texttt{LineItems} which returns an element of the \texttt{Orders} table then it can be done as follows: 

\begin{alltt}
DcColumn order = \\ createColumn("order", "LineItems", "Orders"); 
order.definition = \\ \textbf{"TUPLE( supp=[supplierId], no=[orderNo] )"}; 
\end{alltt}

\noindent
Here the output table \texttt{Orders} is supposed to have columns \texttt{supp} and \texttt{no} while the input table has columns \texttt{supplierId} and \texttt{orderNo}. Once this column has been defined, it can be used in formulas to access orders given line items but a separate join table is never built.

Another mechanism where DataCommander changes the way data is being manipulated is data aggregation. The standard approach assumes that all the data is stored in one table and then it is necessary to specify which columns are used for grouping and aggregation. The result is then returned as a new table with all groups and the corresponding aggregated values for the group members. DataCommandr assumes that one or more aggregated columns can be added to any existing table by providing the corresponding definition. Instead of defining a new table as a result of some set operation (like group-by) we simply define a new \textit{accumulation column} which is supposed to "accumulate" multiple values provided by another table. For example, we can compute the total amount for each order by summing up the prices of all its line items: 

\begin{alltt}
DcColumn total = \\ createColumn("total", "Orders", "Double"); 
total.definition = \\ \textbf{"ACCU( [LineItems], [order], [amount], SUM )"}; 
\end{alltt}

\noindent
Here the accumulation operator \texttt{ACCU} takes four parameters: a fact table (\texttt{LineItems}), a grouping column of the fact table (\texttt{order}), a measure column of the fact table (\texttt{amount}) and an aggregation function (\texttt{SUM}). The procedure will loop through all facts in the table \texttt{LineItems}, retrieve the measure value returned by the \texttt{amount} calculated column, and then add this value to the element pointed to by the \texttt{order} link column. Here we again see that we essentially define a new column in terms of other columns without any set operations. It is much more flexible and intuitive approach especially taking into account that the group, measure and aggregation function parameters can be arbitrary COEL expressions so that we can collect data using rather complex paths between tables. 

\section{Conclusion}
\label{conclusion}

In summary, COM is based on the following basic tenets that distinguish it from other models: 

\textit{Duality principle.} COM emphasizes the differences between what is passed by-value and what is represented indirectly by using other values. Therefore, a data element in COM is defined as a value (identity tuple) with associated other values (entity tuple). Such couples are modeled by a novel type modeling construct, concept, which generalizes conventional classes. A model is then split into two branches -- identity modeling and entity modeling -- by producing a nice yin-yang style of balance and symmetry between two sides of the problem domain. 

\textit{Inclusion principle.} One identity dimension of a data element has a special semantic interpretation. It is supposed to represent a container where this element is a member, a more general element made more specific by this element, and an address space where this element exists by providing an extending segment for it, a base element the properties of which this element inherits. An important difference from other models is that inclusion principle in COM unifies containment (IS-IN) and specialization with inheritance (IS-A). This principle underlies a new function overriding mechanism where a more specific function provides an extension for its base function (rather than completely overriding it). Inclusion principle eliminates the asymmetry between classes and their instances which now both exist in a hierarchy.

\textit{Partial order.} COM assumes that all elements are partially ordered. The main purpose and advantage of partial order is that it can describe many existing semantic relationships: object-attribute-value (object is a lesser element and value is a greater element), multidimensional space (point is a lesser element and coordinate is a greater element), containment (greater elements are collections for lesser elements), relationships (lesser elements relate greater elements). Data modeling is then reduced to partially ordering a set of concepts while their semantics is derived from this structure. 


\end{document}